\documentclass[prd,superscriptaddress,amsmath,amssymb,twocolumn,floatfix]{revtex4}

\usepackage{setspace}
\usepackage{graphicx}
\usepackage{hyperref}

\usepackage{xcolor}

\usepackage{subfigure}
\usepackage{diagbox}
\usepackage{slashbox}
\usepackage{bm}
\usepackage{natbib}
\usepackage{tabularx}
\usepackage{ulem}

\begin{document}

\begin{flushleft}
KCL-PH-TH-2021-69
\end{flushleft}

\title{
 Ability of LISA to detect a gravitational-wave background of cosmological origin: The cosmic string case
}%

\author{Guillaume Boileau}
\affiliation{Artemis, Observatoire de la Côte d'Azur, Université Côte d'Azur, CNRS, 06304 Nice, France}

\author{Alexander C. Jenkins}
\altaffiliation[Present address: ]{Department of Physics and Astronomy, University College London, London WC1E 6BT, United Kingdom}
\affiliation{Theoretical Particle Physics and Cosmology Group, Physics Department,\\ King's College London, University of London, Strand, London WC2R 2LS, United Kingdom}

\author{Mairi Sakellariadou}
\affiliation{Theoretical Particle Physics and Cosmology Group, Physics Department,\\ King's College London, University of London, Strand, London WC2R 2LS, United Kingdom}

\author{Renate Meyer}
\affiliation{Department of Statistics, University of Auckland, Auckland, New Zealand}

\author{Nelson Christensen}
\affiliation{Artemis, Observatoire de la Côte d'Azur, Université Côte d'Azur, CNRS, 06304 Nice, France}


\date{\today}
\begin{abstract}
We investigate the ability of the Laser Interferometer Space Antenna (LISA) to detect a stochastic gravitational-wave background (GWB) produced by cosmic strings, and to subsequently estimate the string tension $G\mu$ in the presence of instrument noise, an astrophysical background from compact binaries, and the galactic foreground from white dwarf binaries. 
Fisher Information and Markov Chain Monte Carlo methods provide estimates of the LISA noise and the parameters for the different signal sources. We demonstrate the importance of including the galactic foreground as well as the astrophysical background for LISA to detect a cosmic string produced GWB and estimate the string tension. Considering the expected astrophysical background and a galactic foreground, a cosmic string tension in the $G \mu \approx 10^{-16}$ to $G\mu \approx 10^{-15}$ range or bigger could be measured by LISA,
with the galactic foreground affecting this limit more than the astrophysical background. The parameter estimation methods presented here can be applied to other cosmological backgrounds in the LISA observation band. 
\end{abstract}
\maketitle

\section{Introduction}%
The Laser Interferometer Space Antenna (LISA)~\cite{2017arXiv170200786A} 
will be able to simultaneously observe a  gravitational-wave  background (GWB) produced by the superposition of
 a large number of independent sources of either cosmological or astrophysical origin~\cite{Christensen_2018}.
 In the LISA frequency band,  [$10^{-5}$, $1$] Hz, we can distinguish different independent sources. The first component is the galactic foreground, from double white dwarf binaries (DWD) in our galaxy~\cite{10.1093/mnras/stz2834}, which will be observed as a modulated waveform~\cite{2014PhRvD..89b2001A}. The second component is the astrophysical background produced by binary black holes (BBH) and binary neutron stars (BNS), which can be predicted from the LIGO/Virgo detections. The energy density of the astrophysical GWB has some uncertainty; the estimation from the LIGO/Virgo detections gives~\cite{2019ApJ...871...97C} $\Omega_{\rm GW}(25 \text{ Hz}) \simeq 1.8 \times 10^{-9} - 2.5 \times 10^{-9}$, while simulations of BBH and BNS populations~\cite{PhysRevD.103.043002} lead to $\Omega_{\rm GW}(25 \text{ Hz})  \simeq 4.97 \times 10^{-9} - 2.58 \times 10^{-8}$. The third component is the cosmological stochastic GWB. The band $10^{-5}$ to $10^{-4}$ Hz, if usable with LISA data, would be important for the detection and separation of the GWB with $A, E, T$ channels. The galactic foreground dominates this low-frequency band. The high frequency (up to $0.1$ Hz) is dominated by the LISA noise, so it is difficult to separate a GWB from this noise. However, the high-frequency data does provide important information about the LISA noise.
 
Given the ability of LISA to simultaneously detect a large number of galactic DWD and a large number of compact binaries, one may question how these contributions may affect the detectability of a GWB of cosmological origin. Here we consider cosmic strings.
Cosmic strings are one of the most promising ways of using gravitational-wave observations to search for physics beyond the standard model, and are thus one of the key cosmological targets of LISA and other experiments. It is therefore vital to understand LISA's sensitivity to a cosmic string GWB signal in the presence of other confusion signals.
Our method can be applied to other cosmological sources, for instance first order phase transitions.

 Cosmic strings~\cite{Vilenkin:2000jqa} are one-dimensional topological defects that may be formed in the  early universe after a phase transition followed by spontaneously broken symmetries. Such objects, relics of a previous more symmetric phase of the universe, appear generically in the context of grand unified theories~\cite{Jeannerot:2003qv}. The authors of~\cite{Auclair_2020} 
 presented a cosmic string stochastic GWB search by LISA in the presence of detector noise. Considering different cosmic string models for the loop distribution~\cite{Kibble:1984hp,Blanco-Pillado:2013qja,Ringeval:2005kr,Lorenz:2010sm}, it was shown that LISA will be able to probe cosmic strings with tensions $G\mu \gtrsim {\cal O}(10^{-17})$, improving by about 6 orders of magnitude current pulsar timing arrays (PTA) constraints, and potentially 3 orders of magnitude with respect to expected constraints from next generation PTA observatories. Advanced LIGO and Advanced Virgo have recently used the data from their first three observing runs to constrain cosmic strings; in particular for one loop distribution model they set the most competitive constraints to date, namely allowing $G\mu \lesssim 4 \times 10^{-15}$~\cite{LIGOScientific:2021nrg}.

In what follows we investigate whether the conclusions of~\cite{Auclair_2020} are affected by considering the contribution from the astrophysical background and the galactic foreground.
We introduce the energy spectral density $\Omega_{\rm GW,G\mu}(G\mu,{\cal M}_i,f)$ from cosmic string loop distributions ${\cal M}_i$ (with $i=1$ for model~\cite{KIBBLE1985227}, $i=2$ for model~\cite{Blanco-Pillado:2013qja}, and $i=3$ for model~\cite{Ringeval:2005kr, Lorenz:2010sm}), at frequency $f$. The total GWB is the sum:  
\begin{equation}\label{eq:SGWB}
\begin{split}
      \Omega_{\rm GW}(f) &= \frac{A_1 \left(\frac{f}{f_*}\right)^{\alpha_1}}{1 + A_2 \left(\frac{f}{f_*}\right)^{\alpha_2}} + \Omega_{\rm astro} \left(\frac{f}{f_*}\right)^{\alpha_{\rm astro}}\\
      &+ \Omega_{{\rm GW},G\mu}(G\mu,{\cal M}_i,f)~,
\end{split}
\end{equation}
 In Eq.~(\ref{eq:SGWB}) above we model the DWD foreground as a broken power law, with  parameters $A_1, A_2, \alpha_1, \alpha_2$, since at high-frequencies ($\simeq 0.1$ Hz) the number of DWDs decreases~\cite{2021arXiv210504283B}. 
We use a model for the BBH/BNS produced GWB based on the stellar mass black hole observations of LIGO and Virgo~\cite{2019ApJ...871...97C}. We simulate an astronomical GWB with 
$\Omega_{\rm astro} = 4.4 \times 10^{-12}$, $\alpha_{\rm astro} = 2/3$ (the predicted slope~\cite{Farmer:2003pa})
and $f_{*} = 3$ mHz~\cite{2019ApJ...871...97C}.
We also simulate galactic DWD  gravitational-wave emission based on different distributions~\cite{Robson:2018ifk,Nelemans:2004qz,10.1093/mnras/stz2834}, and we account for a modulated DWD signal due to the LISA orbital motion~\cite{2014PhRvD..89b2001A}.

\section{Simulation and Estimation}%
We employ the  same parameter estimation methods as in~\cite{2021arXiv210504283B}.
However, instead of a power law representation for the cosmological GWB, we use the cosmic string generated GWB studied in~\cite{Auclair_2020}.
The galactic DWD foreground is approximated by an analytic model, namely the galactic confusion noise; we refer the reader to  Eq.~(14) of~\cite{Robson:2018ifk},  based on model~\cite{Nelemans:2004qz}.
In addition, we study the galactic DWD model of~\cite{10.1093/mnras/stz2834} and use the corresponding estimated GWB parameters~\cite{2021arXiv210504283B}.

We use the LISA time delay interferometry (TDI) $T$ channel to help estimate the LISA noise parameters, while the gravitational-wave signals are in the $A$ and $E$ channels~\cite{PhysRevD.100.104055}. The LISA noise spectrum is modeled in terms of the acceleration noise, $N_{\rm acc} = 1.44 \times 10^{-48} \ \text{s}^{-4} \text{Hz}^{-1}$, and the optical metrology system noise, $N_{\rm pos} = 3.6 \times 10^{-41} \ \text{Hz}^{-1}$~\cite{2017arXiv170200786A}; we refer the reader to~\cite{2021arXiv210504283B} for details. We simultaneously fit the LISA noise parameters in channels $A, E, T$, and we assume that the $T$ channel contains only noise. 
The low frequency band ($10^{-5}$ Hz to $10^{-4}$ Hz) helps to improve our estimation of the galactic foreground.
We use a LISA observation time of four years of continuous data~\cite{Seoane:2021kkk}.

For the cosmic string generated GWB we create a discrete library of dimensionless energy densities $\Omega_{\rm GW,G\mu}(G\mu,{\cal M}_i,f)$~\cite{Jenkins:2018nty}; this GWB must be calculated numerically, and is computationally expensive as a part of the MCMC. We generate 1000 $\Omega_{\rm GW,G\mu}(G\mu,{\cal M}_i,f)$ for $G\mu$ between $1 \times 10^{-22}$ and $1 \times 10^{-8}$ for each model ${\cal M}_i$. The MCMC calls on this library as the tension parameter $G\mu$ is varied. For the size of our library the minimal resolution for the string tension is $\Delta (G\mu)/(G\mu) = 0.033$.  

We consider three cases for the constituents that make up the observed LISA signal. Case (I): LISA noise and a cosmic string generated GWB (Models ${\cal M}_i$ with $i=1, 2, 3)$. There are three parameters 
$\theta = [N_{\rm acc},$ $N_{\rm pos},$ $G\mu]$, and corresponds to~\cite{Auclair_2020}.
Case (II): LISA noise, the galactic foreground, and the astrophysical GWB. There are eight parameters $\theta = [N_{\rm acc},$ $N_{\rm pos},$ $\Omega_{\rm astro},$ $\alpha_{\rm astro},$ $G\mu,$ $A_1,$ $\alpha_1,$ $A_2,$ $\alpha_2 ]$. Case (III): same as case (II) but with the addition of a cosmic string produced GWB, again for the three cosmic string models.
In this case there are nine parameters $\theta = [N_{\rm acc},$ $N_{\rm pos},$ $\Omega_{\rm astro},$ $\alpha_{\rm astro},$ $G\mu,$ $A_1,$ $\alpha_1,$ $A_2,$ $\alpha_2 ]$. 

 The log-likelihood function ($\textbf{d}$ = frequency domain data for the LISA $A,E,T$ TDI channels, $\theta$ = parameters, frequency bin $k$) is given by
 \begin{equation} \label{eq:likelihood}
\begin{split}
     \mathcal{L}(\textbf{d}|\theta) &=  -\frac{1}{2} \sum_{k=0}^N \Bigg[ \frac{d_{A,k}^2}{S_A(f_k)+N_A(f_k)} \\& + \frac{d_{E,k}^2}{S_E(f_k)+N_E(f_k)} + \frac{d_{T,k}^2}{N_T(f_k)} \\&+  \ln\Big(8\pi^3 (S_A(f_k)+NA(f_k))(S_E(f_k)\\&+N_E(f_k))N_T(f_k) \Big) \Bigg] 
\end{split}
\end{equation}
The noise components $N_A(f)=N_E(f)$ and $N_T(f)$ can be written as~\cite{PhysRevD.100.104055}:
\begin{equation}
\left\{
\begin{array}{l}
    N_A = N_1 - N_2, \\
    N_T = N_1 + 2 N_2 
\end{array}
\right.
\end{equation}
with 
\begin{equation}
\left\{
\begin{array}{l}
    N_1(f) = \left(4 S_s(f) + 8\left( 1 + \cos^2\left(\frac{f}{f_{ref}}\right)\right) S_a(f)\right)|W(f)|^2 \\
    N_2(f) = -\left(2 S_s(f) + 8 S_a(f)\right)\cos\left(\frac{f}{f_{ref}}\right)|W(f)|^2
\end{array}
\right.
\end{equation}
$W(f) = 1 - e^{-\frac{2i f}{f_{ref}}}$, $f_{ref} = \frac{2\pi L}{c}$ with LISA arms $L = 2.5\times 10^9 $ m, and 
\begin{equation}
\left\{
\begin{array}{l}
    S_s(f) = N_{Pos} \\
    S_a(f) = \frac{N_{acc}}{(2 \pi f)^4}\left( 1 + \left(\frac{4 \times 10^{-4} \ \text{Hz}}{f} \right)^2 \right).
\end{array}
\right.
\end{equation}

 We use log uniform priors between $-5$ and $5$ for the two LISA noise magnitudes, the three GWB amplitude parameters, and the string tension, $(N_{\rm acc},N_{\rm pos},A_1, A_2,\Omega_{\rm astro},G\mu)$. Uniform priors are used for the slopes $(\alpha_1,\alpha_2, \alpha_{\rm astro})$, between $-1$ and $1$.
 Using Bayes' theorem, we then obtain the posterior distribution of the parameters $p(\theta|\textbf{d}) \propto p(\theta) \mathcal{L}(\textbf{d}|\theta)$ and use a sampling-based approach to posterior computation.
 See \cite{2021arXiv210504283B} for a comprehensive description of these parameter estimation methods and the LISA data.
 
 For parameter estimation, we generate results by two methods. First, we use an adaptive Markov chain Monte Carlo (MCMC)~\cite{doi:10.1198/jcgs.2009.06134}
following a
Metropolis-Hastings algorithm~\cite{10.1093/biomet/57.1.97,gilks1995markov}.
The proposal density $Q_n(x)$ is used to draw candidate parameters. We use this proposal distribution to increase the acceptance rate:
\begin{equation}
    Q_n(x)= (1-\beta)N(x,(2.28)^2 \Sigma_n / d ) + \beta N(x,(0.1)^2 I_d/d)~.
    \label{dens}
\end{equation}
It depends on the number of parameters, $d$, with
$\Sigma_n$ the current empirical estimate of the covariance matrix from samples of the Markov chains, $\beta = 0.25$, 
 $N$ a multinormal distribution, and $I_d$ the identity metric; we refer the reader to~\cite{PhysRevD.103.103529,2021arXiv210504283B}.
 
Alternatively, we use the Fisher Information  $F_{ab}$ to estimate the parameter uncertainties, $\sqrt{F_{aa}^{-1}} = \sigma_a$, as
\begin{equation} \label{eq:Fisher}
\begin{split}
            F_{ab} &= \frac{1}{2} \mathrm{Tr}\left(\mathcal{C}^{-1}\frac{\partial \mathcal{C}} {\partial \theta_a} \mathcal{C}^{-1} \frac{\partial \mathcal{C}}{\partial\theta_b}  \right) \\&=  \sum_{I=A,E,T} \sum_{k=0}^N \frac{M}{2} \frac{\frac{\partial S_I(f_k)+ N_I(f_k)} {\partial \theta_a}\frac{\partial S_I(f_k)+ N_I(f_k)} {\partial \theta_b}}{\left(S_I(f_k)+ N_I(f_k)\right)^2}~.
\end{split}
\end{equation}
with $M=Df_b$ where $D$ is the time duration of the LISA mission and $f_b$ the highest frequency of interest in the LISA band~\cite{PhysRevD.100.104055}. 
The correlation matrix is 
\begin{equation}\label{eq:COV}
    \mathcal{C}(\theta,f) =
     \left(
     \begin{array}{ccc}
      S_A + N_A & 0 & 0  \\
      0 & S_E + N_E & 0 \\
      0 & 0 & N_T, \\
     \end{array}
     \right)~,
   \end{equation}
where $N_I$, with $I = [A,E,T]$, are the noise power spectral densities for the different TDI channels. The signal channels ($A,E$) power spectral densities in (\ref{eq:Fisher}) are  
\begin{equation}\label{eq:signal_psd}
S_I(f) = \frac{3H_0^2}{4 \pi^2} \frac{ \sum_i \Omega_{{\rm GW},i}}{\mathcal{R}(f)f^3} ~.
\end{equation}

We compare signals with and without a cosmic string produced GWB to study detectability and parameter estimation accuracy in the presence of a galactic foreground and the astrophysical GWB. We do not do model comparison between the different cosmic string models; this will be the subject of a future study.
With the output chains from the MCMC analyses we use the deviance information criterion for model comparison~\cite{https://doi.org/10.1111/1467-9868.00353,10.2307/24774528,doi:https://doi.org/10.1002/9781118445112.stat07878}. 
Note that the use of improper priors precludes the use of Bayes factors for model comparison.

\section{Results}
\begin{figure*}
\includegraphics[width=\linewidth]{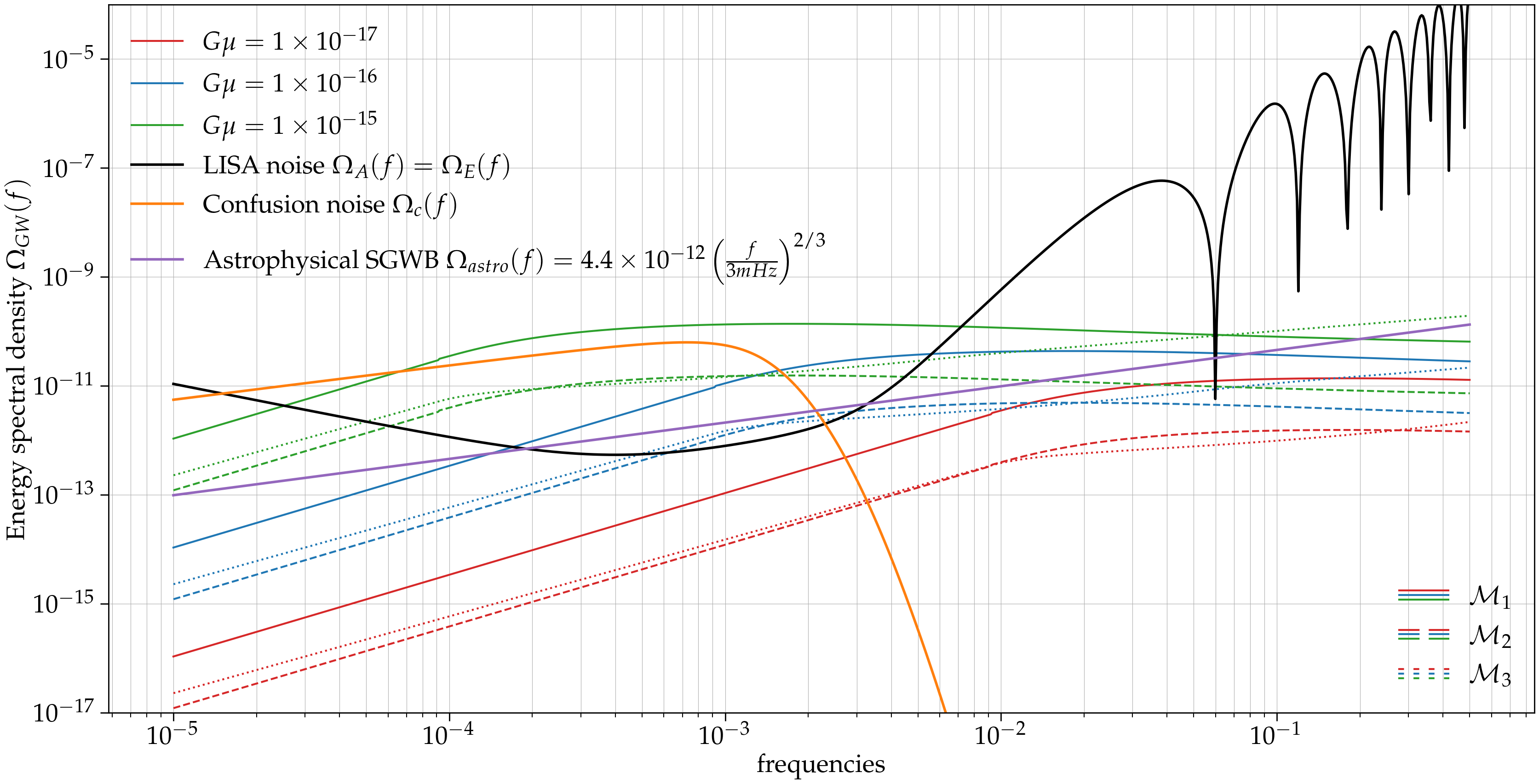}
\caption{Dimensionless energy densities $\Omega_{\rm GW,G\mu}(f)$ for cosmic string models ${\cal M}_i$ with $i=1,2,3$~\cite{Auclair_2020}, displayed respectively with lines, long dashes and short dashes. The red, blue and green colors correspond to the cosmic sting GWBs for three inputs, respectively, $G\mu = 1\times 10^{-17}$, $1\times 10^{-18}$ and $1\times 10^{-19} $. We include LISA noise~\cite{PhysRevD.100.104055} for four years of observation: $\Omega_A(f) = \Omega_E(f)$ (black line), a galactic confusion noise~\cite{Robson:2018ifk} $ \Omega_c(f)$ (orange line), and an astrophysical compact binary background~\cite{2019ApJ...871...97C}, $\Omega_{astro}(f) = 4.4 \times 10^{-12} \left(\frac{f}{3 mHz} \right)^{2/3}$ (purple line).}
\label{fig:model}
\end{figure*}

We display in Fig.~\ref{fig:model}  the dimensionless GWB energy density for the three cosmic string models, and for string tension values around the level of LISA detectability~\cite{Jenkins:2018nty}. We also show the LISA noise~\cite{PhysRevD.100.104055}, a galactic confusion noise model~\cite{Robson:2018ifk}, and an astrophysical compact binary background~\cite{2019ApJ...871...97C}. 

\begin{figure*}
\includegraphics[width=\linewidth]{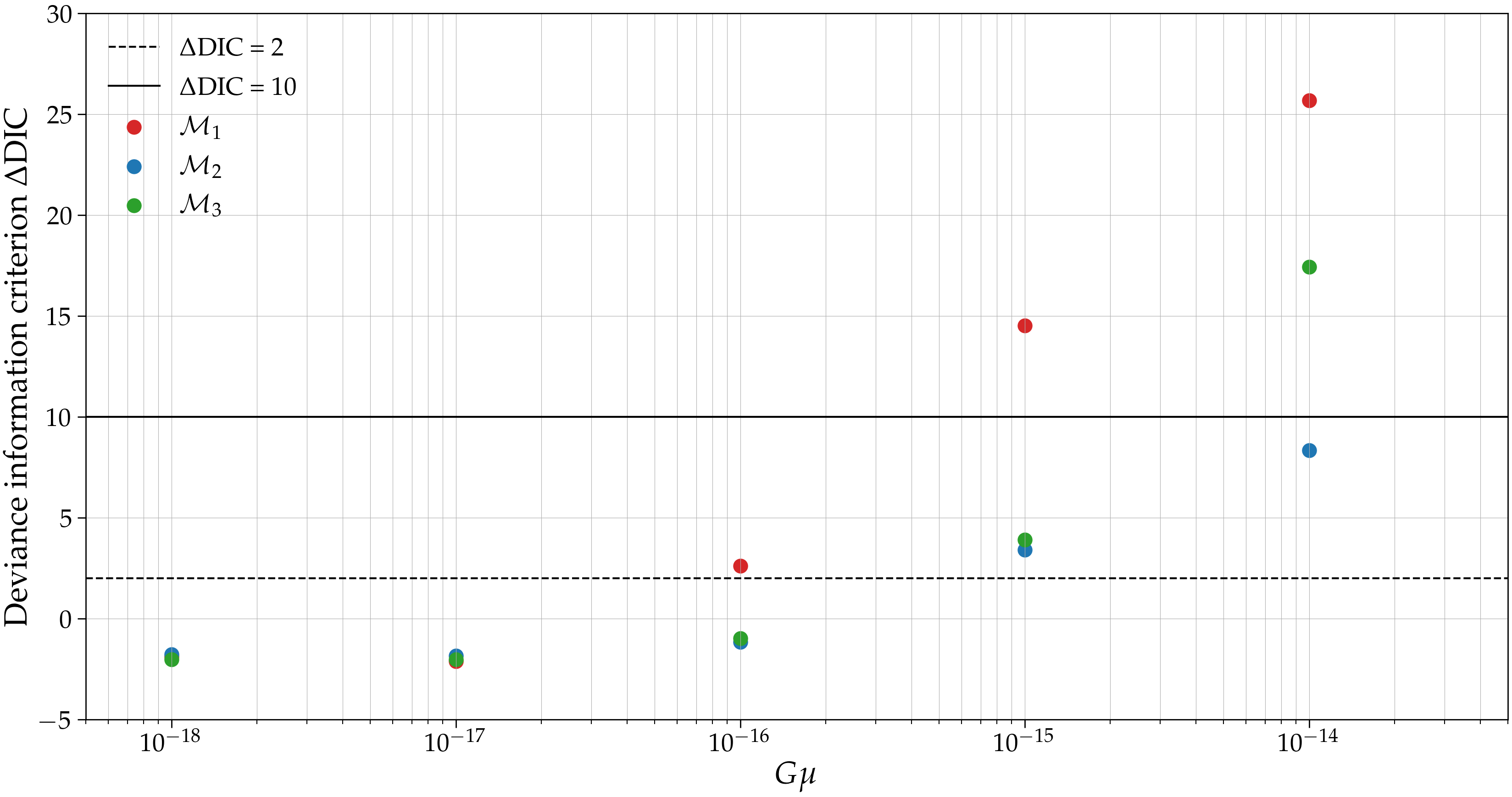}
\caption{Model comparison between cases (II) and (III) using the Deviance information criterion. 
Strong evidence for the presence of the cosmic string signal begins in the $G\mu \approx 10^{-16}$ (${\cal M}_1$) to $G\mu \approx 10^{-15}$ (${\cal M}_2$ and ${\cal M}_3$) range for the cosmic string models.}
\label{fig:bic1}
\end{figure*}

The ability to observe a cosmic string produced GWB is displayed in Fig.~\ref{fig:bic1} using the Deviance information criterion. For each cosmic string model ${\cal M}_i$ we compare cases (II) and (III).

Similar to the Jeffreys' scale, the difference in the Deviance information criterion ($\Delta {\rm DIC}$) starts to give  evidence for a particular model for 
$\Delta {\rm DIC}>2$, is substantial for $\Delta {\rm DIC}$ between 5 and 10,
and becomes decisive for $\Delta {\rm DIC} > 10$~\cite{doi:10.1080/01621459.1995.10476572,LunnDavid2013TBb:}; see \cite{Anagnostopoulos:2019miu} for another use of the DIC in physics. Hence, the ability for LISA to detect a cosmic string produced GWB begins in the $G\mu \approx 10^{-16}$ (${\cal M}_1$) to $G\mu \approx 10^{-15}$ (${\cal M}_2$ and ${\cal M}_3$) range for all three cosmic string models.

\begin{figure*}
\includegraphics[width=\linewidth]{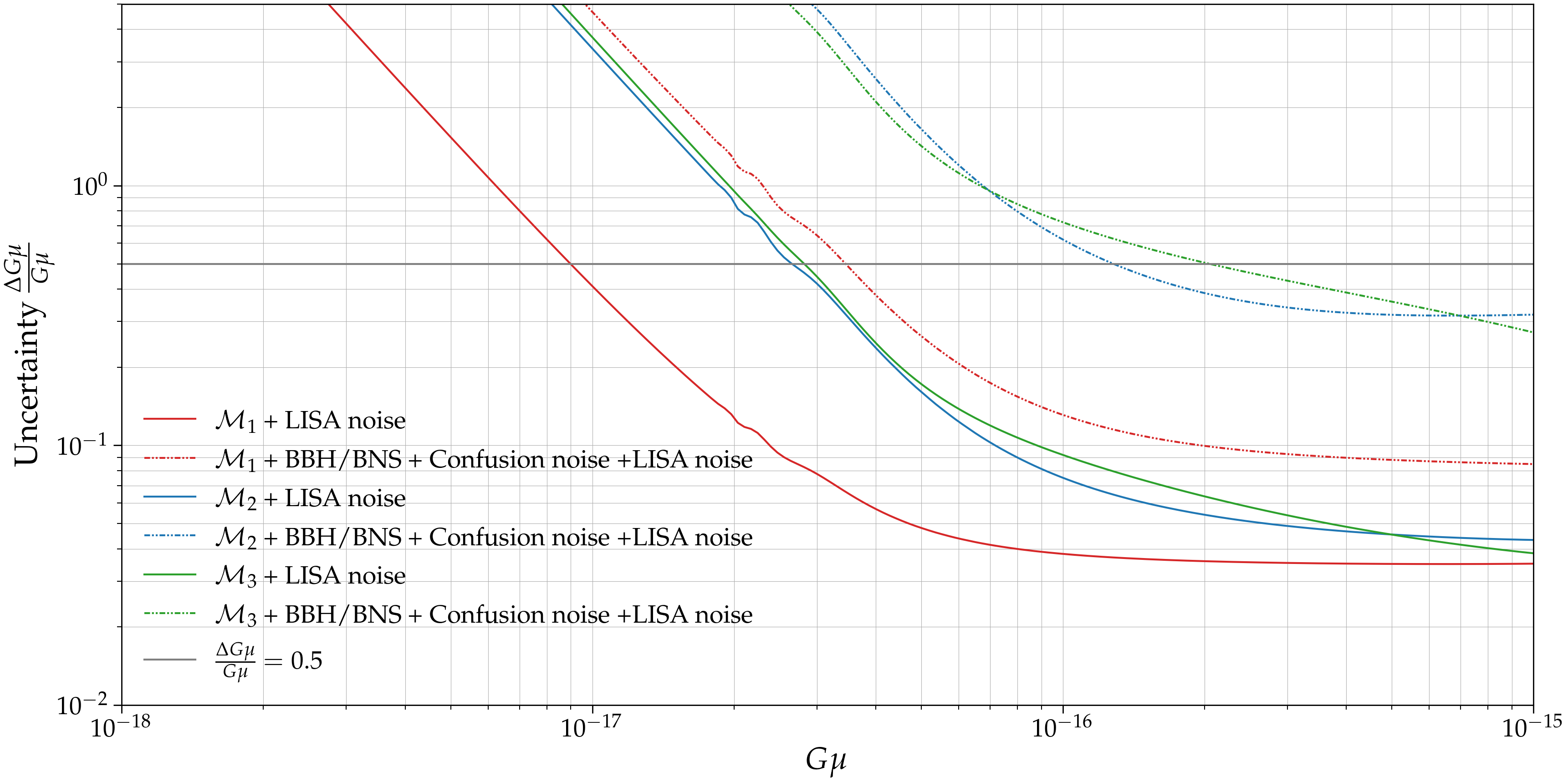}
\caption{Cosmic string tension $G \mu$ uncertainty estimates from the Fisher information study for the three models ${\cal M}_i$. Solid lines present the results for 
case (I).
The dot-dashed lines are the results considering 
case (III).
The horizontal gray line represents $\Delta G\mu/G\mu = 0.5$. A MCMC study gives equivalent results.}
\label{fig:Uncertaintytotal}
\end{figure*}

We present in Fig.~\ref{fig:Uncertaintytotal} the $G\mu$ uncertainty estimates following the Fisher information analysis. The adaptive MCMC produces equivalent results, as was the case in~\cite{PhysRevD.103.103529,2021arXiv210504283B}, but for simplicity in Fig.~\ref{fig:Uncertaintytotal} we only show the Fisher information results.
 We conduct the study with different cosmic string generated GWBs by changing the string tension $G\mu$.
 
 The horizontal dashed line represents the accuracy level of $50\%$; above this limit we cannot separate the cosmic strings GWB.  We display the results for cases (I) and (III), showing that the inclusion of the galactic foreground and astrophysical background has an important effect on the parameter estimation for the cosmic string tension.
 We summarize the ability to conduct parameter estimation on the cosmic string tension for cases (I) and (III) in Table~\ref{tab:result}.
We also performed a study 
following the method~\cite{10.1093/mnras/stz2834} 
as in~\cite{2021arXiv210504283B}. We obtained similar results to the ones presented in Table~\ref{tab:result}.
In addition, we have repeated the Fisher information study presented in Fig.~\ref{fig:Uncertaintytotal} but with the reduced bandwidth of $10^{-4}$ Hz to 0.1 Hz, and the conclusions do not change.

\begin{table*}[htbp]
 \begin{tabular}{ lccc}
 \hline 
   \multicolumn{4}{c}{\textbf{LISA noise + Cosmic strings}}\\
 \hline 
  & ${\cal M}_1$ & ${\cal M}_2$   & ${\cal M}_3$  \\ \hline 
$G\mu_{lim}$ & $1\times 10^{-17} $   &$3\times 10^{-17} $   & $3\times 10^{-17}  $  \\  \hline
\multicolumn{4}{c}{\textbf{LISA noise + DWD + BBH/BNS + Cosmic strings}}\\
 \hline 
  & ${\cal M}_1$ & ${\cal M}_2$   & ${\cal M}_3$  \\ \hline 
$G\mu_{lim}$ & $3 \times 10^{-17} $   &$1\times 10^{-16} $   & $2\times 10^{-16}  $  \\    \hline
\end{tabular}
 \caption{Table summarizing the results of the ability to observe a GWB from cosmic strings (the three models ${\cal M}_i$) in the presence of just LISA noise, or with LISA noise, the DWD produced galactic foreground and a compact binary astrophysical background. We report the values where the uncertainty in the estimation of the string tension is $\Delta (G\mu)/(G\mu) = 0.5$. We display in Fig.~\ref{fig:Uncertaintytotal} the way the uncertainty diminishes as $G\mu$ increases.}
 \label{tab:result}
\end{table*} 

The sum of galactic foreground and astrophysical background have an effect on the detection of a cosmic string produced GWB and the estimation of the string tension $G\mu$. In fact, the galactic foreground is the most important for limiting the cosmic string measurement as it is dominant at low frequencies; see Fig.~\ref{fig:model}. 

\section{Conclusion}
We have studied the LISA measurement limit of a cosmological GWB from cosmic strings.
This cosmic string GWB is in the presence of a compact binary produced astrophysical background, a galactic foreground, and LISA noise for four years of data. The detection limit for the three loop distribution models (studied in~\cite{Auclair_2020}) is in the $G\mu \approx 10^{-16}$ (${\cal M}_1$) to $G\mu \approx 10^{-15}$ (${\cal M}_2$ and ${\cal M}_3$) range for four years of LISA observations. The ability to conduct parameter estimation and resolve the value of the cosmic string tension begins around $G\mu \approx 10^{-16}$. Certainly if other cosmological GWBs are present, for example first order phase transitions or inflation~\cite{Caprini:2018mtu}, then even more sophisticated parameter estimation methods will be needed. 
Principal component analysis has been proposed for making such a cosmological GWB detection, but does not address parameter estimation~\cite{Pieroni:2020rob}.
Note too that for the LISA GWB search to be most effective a more comprehensive description will need to be made for the LISA noise, plus the incorporation of methods that can observe and parameterize individually resolvable binary systems in the galaxy~\cite{2014PhRvD..89b2001A,Babak_2017}, as well as other expected, overlapping signals in the LISA band (massive black hole binaries, extreme mass ratio inspirals, etc.)~\cite{2017arXiv170200786A}. It is important to note that the results presented here are obtained with a simple LISA noise model generated with two parameters~\cite{PhysRevD.100.104055}. Additional low frequency noise would make the detection less efficient. More complex LISA noise models will certainly need to be investigated. These issues will be addressed in future studies.

\section{Acknowledgments}
G.B. and N.C. thank the Centre national d'\'etudes spatiales for support for this research.  A.C.J. was supported by King's College London through a Graduate Teaching Scholarship. M.S. was supported in part by the Science and Technology Facility Council (STFC), United Kingdom, under the research grant No. ST/P000258/1. MS thanks the laboratory  Artemis, Observatoire de la C\^ote d'Azur, for hospitality. R.M. acknowledges support by the James Cook Research  Fellowship JCRF-UOA-1801 from Government funding, administered by the Royal Society Te Aparangi, and DFG Grant  No. KI 1443/3-2.

\bibliography{Biblio}

\begin{thebibliography}{35}
\expandafter\ifx\csname natexlab\endcsname\relax\def\natexlab#1{#1}\fi
\expandafter\ifx\csname bibnamefont\endcsname\relax
  \def\bibnamefont#1{#1}\fi
\expandafter\ifx\csname bibfnamefont\endcsname\relax
  \def\bibfnamefont#1{#1}\fi
\expandafter\ifx\csname citenamefont\endcsname\relax
  \def\citenamefont#1{#1}\fi
\expandafter\ifx\csname url\endcsname\relax
  \def\url#1{\texttt{#1}}\fi
\expandafter\ifx\csname urlprefix\endcsname\relax\def\urlprefix{URL }\fi
\providecommand{\bibinfo}[2]{#2}
\providecommand{\eprint}[2][]{\url{#2}}

\bibitem[{\citenamefont{{Amaro-Seoane} et~al.}(2017)}]{2017arXiv170200786A}
\bibinfo{author}{\bibfnamefont{P.}~\bibnamefont{{Amaro-Seoane}}}
  \bibnamefont{et~al.}, \bibinfo{journal}{arXiv e-prints}
  \bibinfo{eid}{arXiv:1702.00786} (\bibinfo{year}{2017}), \eprint{1702.00786}.

\bibitem[{\citenamefont{Christensen}(2019)}]{Christensen_2018}
\bibinfo{author}{\bibfnamefont{N.}~\bibnamefont{Christensen}},
  \bibinfo{journal}{Reports on Progress in Physics}
  \textbf{\bibinfo{volume}{82}}, \bibinfo{pages}{016903}
  (\bibinfo{year}{2019}), ISSN \bibinfo{issn}{1361-6633},
  \urlprefix\url{http://dx.doi.org/10.1088/1361-6633/aae6b5}.

\bibitem[{\citenamefont{Lamberts et~al.}(2019)\citenamefont{Lamberts, Blunt,
  Littenberg, Garrison-Kimmel, Kupfer, and Sanderson}}]{10.1093/mnras/stz2834}
\bibinfo{author}{\bibfnamefont{A.}~\bibnamefont{Lamberts}},
  \bibinfo{author}{\bibfnamefont{S.}~\bibnamefont{Blunt}},
  \bibinfo{author}{\bibfnamefont{T.~B.} \bibnamefont{Littenberg}},
  \bibinfo{author}{\bibfnamefont{S.}~\bibnamefont{Garrison-Kimmel}},
  \bibinfo{author}{\bibfnamefont{T.}~\bibnamefont{Kupfer}}, \bibnamefont{and}
  \bibinfo{author}{\bibfnamefont{R.~E.} \bibnamefont{Sanderson}},
  \bibinfo{journal}{Monthly Notices of the Royal Astronomical Society}
  \textbf{\bibinfo{volume}{490}}, \bibinfo{pages}{5888} (\bibinfo{year}{2019}),
  ISSN \bibinfo{issn}{0035-8711},
  \eprint{http://oup.prod.sis.lan/mnras/article-pdf/490/4/5888/30995029/stz2834.pdf},
  \urlprefix\url{https://doi.org/10.1093/mnras/stz2834}.

\bibitem[{\citenamefont{{Adams} and {Cornish}}(2014)}]{2014PhRvD..89b2001A}
\bibinfo{author}{\bibfnamefont{M.~R.} \bibnamefont{{Adams}}} \bibnamefont{and}
  \bibinfo{author}{\bibfnamefont{N.~J.} \bibnamefont{{Cornish}}},
  \bibinfo{journal}{\prd} \textbf{\bibinfo{volume}{89}}, \bibinfo{eid}{022001}
  (\bibinfo{year}{2014}), \eprint{1307.4116}.

\bibitem[{\citenamefont{{Chen} et~al.}(2019)\citenamefont{{Chen}, {Huang}, and
  {Huang}}}]{2019ApJ...871...97C}
\bibinfo{author}{\bibfnamefont{Z.-C.} \bibnamefont{{Chen}}},
  \bibinfo{author}{\bibfnamefont{F.}~\bibnamefont{{Huang}}}, \bibnamefont{and}
  \bibinfo{author}{\bibfnamefont{Q.-G.} \bibnamefont{{Huang}}},
  \bibinfo{journal}{APJ} \textbf{\bibinfo{volume}{871}}, \bibinfo{eid}{97}
  (\bibinfo{year}{2019}), \eprint{1809.10360}.

\bibitem[{\citenamefont{P\'erigois et~al.}(2021)\citenamefont{P\'erigois,
  Belczynski, Bulik, and Regimbau}}]{PhysRevD.103.043002}
\bibinfo{author}{\bibfnamefont{C.}~\bibnamefont{P\'erigois}},
  \bibinfo{author}{\bibfnamefont{C.}~\bibnamefont{Belczynski}},
  \bibinfo{author}{\bibfnamefont{T.}~\bibnamefont{Bulik}}, \bibnamefont{and}
  \bibinfo{author}{\bibfnamefont{T.}~\bibnamefont{Regimbau}},
  \bibinfo{journal}{Phys. Rev. D} \textbf{\bibinfo{volume}{103}},
  \bibinfo{pages}{043002} (\bibinfo{year}{2021}),
  \urlprefix\url{https://link.aps.org/doi/10.1103/PhysRevD.103.043002}.

\bibitem[{\citenamefont{Vilenkin and Shellard}(2000)}]{Vilenkin:2000jqa}
\bibinfo{author}{\bibfnamefont{A.}~\bibnamefont{Vilenkin}} \bibnamefont{and}
  \bibinfo{author}{\bibfnamefont{E.~P.~S.} \bibnamefont{Shellard}},
  \emph{\bibinfo{title}{{Cosmic Strings and Other Topological Defects}}}
  (\bibinfo{publisher}{Cambridge University Press}, \bibinfo{year}{2000}), ISBN
  \bibinfo{isbn}{978-0-521-65476-0}.

\bibitem[{\citenamefont{Jeannerot et~al.}(2003)\citenamefont{Jeannerot, Rocher,
  and Sakellariadou}}]{Jeannerot:2003qv}
\bibinfo{author}{\bibfnamefont{R.}~\bibnamefont{Jeannerot}},
  \bibinfo{author}{\bibfnamefont{J.}~\bibnamefont{Rocher}}, \bibnamefont{and}
  \bibinfo{author}{\bibfnamefont{M.}~\bibnamefont{Sakellariadou}},
  \bibinfo{journal}{Phys. Rev. D} \textbf{\bibinfo{volume}{68}},
  \bibinfo{pages}{103514} (\bibinfo{year}{2003}), \eprint{hep-ph/0308134}.

\bibitem[{\citenamefont{Auclair et~al.}(2020)\citenamefont{Auclair,
  Blanco-Pillado, Figueroa, Jenkins, Lewicki, Sakellariadou, Sanidas, Sousa,
  Steer, Wachter et~al.}}]{Auclair_2020}
\bibinfo{author}{\bibfnamefont{P.}~\bibnamefont{Auclair}},
  \bibinfo{author}{\bibfnamefont{J.~J.} \bibnamefont{Blanco-Pillado}},
  \bibinfo{author}{\bibfnamefont{D.~G.} \bibnamefont{Figueroa}},
  \bibinfo{author}{\bibfnamefont{A.~C.} \bibnamefont{Jenkins}},
  \bibinfo{author}{\bibfnamefont{M.}~\bibnamefont{Lewicki}},
  \bibinfo{author}{\bibfnamefont{M.}~\bibnamefont{Sakellariadou}},
  \bibinfo{author}{\bibfnamefont{S.}~\bibnamefont{Sanidas}},
  \bibinfo{author}{\bibfnamefont{L.}~\bibnamefont{Sousa}},
  \bibinfo{author}{\bibfnamefont{D.~A.} \bibnamefont{Steer}},
  \bibinfo{author}{\bibfnamefont{J.~M.} \bibnamefont{Wachter}},
  \bibnamefont{et~al.}, \bibinfo{journal}{Journal of Cosmology and
  Astroparticle Physics} \textbf{\bibinfo{volume}{2020}}, \bibinfo{pages}{034}
  (\bibinfo{year}{2020}),
  \urlprefix\url{https://doi.org/10.1088/1475-7516/2020/04/034}.

\bibitem[{\citenamefont{Kibble}(1985{\natexlab{a}})}]{Kibble:1984hp}
\bibinfo{author}{\bibfnamefont{T.~W.~B.} \bibnamefont{Kibble}},
  \bibinfo{journal}{Nucl. Phys. B} \textbf{\bibinfo{volume}{252}},
  \bibinfo{pages}{227} (\bibinfo{year}{1985}{\natexlab{a}}),
  \bibinfo{note}{[Erratum: Nucl.Phys.B 261, 750 (1985)]}.

\bibitem[{\citenamefont{Blanco-Pillado
  et~al.}(2014)\citenamefont{Blanco-Pillado, Olum, and
  Shlaer}}]{Blanco-Pillado:2013qja}
\bibinfo{author}{\bibfnamefont{J.~J.} \bibnamefont{Blanco-Pillado}},
  \bibinfo{author}{\bibfnamefont{K.~D.} \bibnamefont{Olum}}, \bibnamefont{and}
  \bibinfo{author}{\bibfnamefont{B.}~\bibnamefont{Shlaer}},
  \bibinfo{journal}{Phys. Rev. D} \textbf{\bibinfo{volume}{89}},
  \bibinfo{pages}{023512} (\bibinfo{year}{2014}), \eprint{1309.6637}.

\bibitem[{\citenamefont{Ringeval et~al.}(2007)\citenamefont{Ringeval,
  Sakellariadou, and Bouchet}}]{Ringeval:2005kr}
\bibinfo{author}{\bibfnamefont{C.}~\bibnamefont{Ringeval}},
  \bibinfo{author}{\bibfnamefont{M.}~\bibnamefont{Sakellariadou}},
  \bibnamefont{and} \bibinfo{author}{\bibfnamefont{F.}~\bibnamefont{Bouchet}},
  \bibinfo{journal}{JCAP} \textbf{\bibinfo{volume}{02}}, \bibinfo{pages}{023}
  (\bibinfo{year}{2007}), \eprint{astro-ph/0511646}.

\bibitem[{\citenamefont{Lorenz et~al.}(2010)\citenamefont{Lorenz, Ringeval, and
  Sakellariadou}}]{Lorenz:2010sm}
\bibinfo{author}{\bibfnamefont{L.}~\bibnamefont{Lorenz}},
  \bibinfo{author}{\bibfnamefont{C.}~\bibnamefont{Ringeval}}, \bibnamefont{and}
  \bibinfo{author}{\bibfnamefont{M.}~\bibnamefont{Sakellariadou}},
  \bibinfo{journal}{JCAP} \textbf{\bibinfo{volume}{10}}, \bibinfo{pages}{003}
  (\bibinfo{year}{2010}), \eprint{1006.0931}.

\bibitem[{\citenamefont{Abbott et~al.}(2021)}]{LIGOScientific:2021nrg}
\bibinfo{author}{\bibfnamefont{R.}~\bibnamefont{Abbott}} \bibnamefont{et~al.}
  (\bibinfo{collaboration}{LIGO Scientific, Virgo, KAGRA}),
  \bibinfo{journal}{Phys. Rev. Lett.} \textbf{\bibinfo{volume}{126}},
  \bibinfo{pages}{241102} (\bibinfo{year}{2021}), \eprint{2101.12248}.

\bibitem[{\citenamefont{Kibble}(1985{\natexlab{b}})}]{KIBBLE1985227}
\bibinfo{author}{\bibfnamefont{T.}~\bibnamefont{Kibble}},
  \bibinfo{journal}{Nuclear Physics B} \textbf{\bibinfo{volume}{252}},
  \bibinfo{pages}{227} (\bibinfo{year}{1985}{\natexlab{b}}), ISSN
  \bibinfo{issn}{0550-3213},
  \urlprefix\url{https://www.sciencedirect.com/science/article/pii/0550321385904390}.

\bibitem[{\citenamefont{Boileau
  et~al.}(2021{\natexlab{a}})\citenamefont{Boileau, Lamberts, Christensen,
  Cornish, and Meyer}}]{2021arXiv210504283B}
\bibinfo{author}{\bibfnamefont{G.}~\bibnamefont{Boileau}},
  \bibinfo{author}{\bibfnamefont{A.}~\bibnamefont{Lamberts}},
  \bibinfo{author}{\bibfnamefont{N.}~\bibnamefont{Christensen}},
  \bibinfo{author}{\bibfnamefont{N.~J.} \bibnamefont{Cornish}},
  \bibnamefont{and} \bibinfo{author}{\bibfnamefont{R.}~\bibnamefont{Meyer}},
  \bibinfo{journal}{Monthly Notices of the Royal Astronomical Society}
  \textbf{\bibinfo{volume}{508}}, \bibinfo{pages}{803}
  (\bibinfo{year}{2021}{\natexlab{a}}),
  \urlprefix\url{https://doi.org/10.1093/mnras/stab2575}.

\bibitem[{\citenamefont{Farmer and Phinney}(2003)}]{Farmer:2003pa}
\bibinfo{author}{\bibfnamefont{A.~J.} \bibnamefont{Farmer}} \bibnamefont{and}
  \bibinfo{author}{\bibfnamefont{E.}~\bibnamefont{Phinney}},
  \bibinfo{journal}{Mon. Not. Roy. Astron. Soc.}
  \textbf{\bibinfo{volume}{346}}, \bibinfo{pages}{1197} (\bibinfo{year}{2003}),
  \eprint{astro-ph/0304393}.

\bibitem[{\citenamefont{Robson et~al.}(2019)\citenamefont{Robson, Cornish, and
  Liu}}]{Robson:2018ifk}
\bibinfo{author}{\bibfnamefont{T.}~\bibnamefont{Robson}},
  \bibinfo{author}{\bibfnamefont{N.~J.} \bibnamefont{Cornish}},
  \bibnamefont{and} \bibinfo{author}{\bibfnamefont{C.}~\bibnamefont{Liu}},
  \bibinfo{journal}{Class. Quant. Grav.} \textbf{\bibinfo{volume}{36}},
  \bibinfo{pages}{105011} (\bibinfo{year}{2019}), \eprint{1803.01944}.

\bibitem[{\citenamefont{Nelemans and Tout}(2005)}]{Nelemans:2004qz}
\bibinfo{author}{\bibfnamefont{G.}~\bibnamefont{Nelemans}} \bibnamefont{and}
  \bibinfo{author}{\bibfnamefont{C.~A.} \bibnamefont{Tout}},
  \bibinfo{journal}{Mon. Not. Roy. Astron. Soc.}
  \textbf{\bibinfo{volume}{356}}, \bibinfo{pages}{753} (\bibinfo{year}{2005}),
  \eprint{astro-ph/0410301}.

\bibitem[{\citenamefont{Smith and Caldwell}(2019)}]{PhysRevD.100.104055}
\bibinfo{author}{\bibfnamefont{T.~L.} \bibnamefont{Smith}} \bibnamefont{and}
  \bibinfo{author}{\bibfnamefont{R.~R.} \bibnamefont{Caldwell}},
  \bibinfo{journal}{Phys. Rev. D} \textbf{\bibinfo{volume}{100}},
  \bibinfo{pages}{104055} (\bibinfo{year}{2019}),
  \urlprefix\url{https://link.aps.org/doi/10.1103/PhysRevD.100.104055}.

\bibitem[{\citenamefont{Seoane et~al.}(2021)}]{Seoane:2021kkk}
\bibinfo{author}{\bibfnamefont{P.~A.} \bibnamefont{Seoane}}
  \bibnamefont{et~al.} (\bibinfo{year}{2021}), \eprint{2107.09665}.

\bibitem[{\citenamefont{Jenkins and Sakellariadou}(2018)}]{Jenkins:2018nty}
\bibinfo{author}{\bibfnamefont{A.~C.} \bibnamefont{Jenkins}} \bibnamefont{and}
  \bibinfo{author}{\bibfnamefont{M.}~\bibnamefont{Sakellariadou}},
  \bibinfo{journal}{Phys. Rev. D} \textbf{\bibinfo{volume}{98}},
  \bibinfo{pages}{063509} (\bibinfo{year}{2018}), \eprint{1802.06046}.

\bibitem[{\citenamefont{Roberts and
  Rosenthal}(2009)}]{doi:10.1198/jcgs.2009.06134}
\bibinfo{author}{\bibfnamefont{G.~O.} \bibnamefont{Roberts}} \bibnamefont{and}
  \bibinfo{author}{\bibfnamefont{J.~S.} \bibnamefont{Rosenthal}},
  \bibinfo{journal}{Journal of Computational and Graphical Statistics}
  \textbf{\bibinfo{volume}{18}}, \bibinfo{pages}{349} (\bibinfo{year}{2009}).

\bibitem[{\citenamefont{Hastings}(1970)}]{10.1093/biomet/57.1.97}
\bibinfo{author}{\bibfnamefont{W.~K.} \bibnamefont{Hastings}},
  \bibinfo{journal}{Biometrika} \textbf{\bibinfo{volume}{57}},
  \bibinfo{pages}{97} (\bibinfo{year}{1970}), ISSN \bibinfo{issn}{0006-3444}.

\bibitem[{\citenamefont{Gilks et~al.}(1995)\citenamefont{Gilks, Richardson, and
  Spiegelhalter}}]{gilks1995markov}
\bibinfo{author}{\bibfnamefont{W.}~\bibnamefont{Gilks}},
  \bibinfo{author}{\bibfnamefont{S.}~\bibnamefont{Richardson}},
  \bibnamefont{and}
  \bibinfo{author}{\bibfnamefont{D.}~\bibnamefont{Spiegelhalter}},
  \emph{\bibinfo{title}{Markov Chain Monte Carlo in Practice}}, Chapman \&
  Hall/CRC Interdisciplinary Statistics (\bibinfo{publisher}{Taylor \&
  Francis}, \bibinfo{year}{1995}), ISBN \bibinfo{isbn}{9780412055515}.

\bibitem[{\citenamefont{Boileau
  et~al.}(2021{\natexlab{b}})\citenamefont{Boileau, Christensen, Meyer, and
  Cornish}}]{PhysRevD.103.103529}
\bibinfo{author}{\bibfnamefont{G.}~\bibnamefont{Boileau}},
  \bibinfo{author}{\bibfnamefont{N.}~\bibnamefont{Christensen}},
  \bibinfo{author}{\bibfnamefont{R.}~\bibnamefont{Meyer}}, \bibnamefont{and}
  \bibinfo{author}{\bibfnamefont{N.~J.} \bibnamefont{Cornish}},
  \bibinfo{journal}{Phys. Rev. D} \textbf{\bibinfo{volume}{103}},
  \bibinfo{pages}{103529} (\bibinfo{year}{2021}{\natexlab{b}}),
  \urlprefix\url{https://link.aps.org/doi/10.1103/PhysRevD.103.103529}.

\bibitem[{\citenamefont{Spiegelhalter et~al.}(2002)\citenamefont{Spiegelhalter,
  Best, Carlin, and Van Der~Linde}}]{https://doi.org/10.1111/1467-9868.00353}
\bibinfo{author}{\bibfnamefont{D.~J.} \bibnamefont{Spiegelhalter}},
  \bibinfo{author}{\bibfnamefont{N.~G.} \bibnamefont{Best}},
  \bibinfo{author}{\bibfnamefont{B.~P.} \bibnamefont{Carlin}},
  \bibnamefont{and} \bibinfo{author}{\bibfnamefont{A.}~\bibnamefont{Van
  Der~Linde}}, \bibinfo{journal}{Journal of the Royal Statistical Society:
  Series B (Statistical Methodology)} \textbf{\bibinfo{volume}{64}},
  \bibinfo{pages}{583} (\bibinfo{year}{2002}),
  \eprint{https://rss.onlinelibrary.wiley.com/doi/pdf/10.1111/1467-9868.00353},
  \urlprefix\url{https://rss.onlinelibrary.wiley.com/doi/abs/10.1111/1467-9868.00353}.

\bibitem[{\citenamefont{Spiegelhalter et~al.}(2014)\citenamefont{Spiegelhalter,
  Best, Carlin, and van~der Linde}}]{10.2307/24774528}
\bibinfo{author}{\bibfnamefont{D.~J.} \bibnamefont{Spiegelhalter}},
  \bibinfo{author}{\bibfnamefont{N.~G.} \bibnamefont{Best}},
  \bibinfo{author}{\bibfnamefont{B.~P.} \bibnamefont{Carlin}},
  \bibnamefont{and} \bibinfo{author}{\bibfnamefont{A.}~\bibnamefont{van~der
  Linde}}, \bibinfo{journal}{Journal of the Royal Statistical Society. Series B
  (Statistical Methodology)} \textbf{\bibinfo{volume}{76}},
  \bibinfo{pages}{485} (\bibinfo{year}{2014}), ISSN \bibinfo{issn}{13697412,
  14679868}.

\bibitem[{\citenamefont{Meyer}(2016)}]{doi:https://doi.org/10.1002/9781118445112.stat07878}
\bibinfo{author}{\bibfnamefont{R.}~\bibnamefont{Meyer}},
  \emph{\bibinfo{title}{Deviance Information Criterion (DIC)}}
  (\bibinfo{publisher}{Wiley \& Sons}, \bibinfo{year}{2016}), pp.
  \bibinfo{pages}{1--6}, ISBN \bibinfo{isbn}{9781118445112}.

\bibitem[{\citenamefont{Kass and
  Raftery}(1995)}]{doi:10.1080/01621459.1995.10476572}
\bibinfo{author}{\bibfnamefont{R.~E.} \bibnamefont{Kass}} \bibnamefont{and}
  \bibinfo{author}{\bibfnamefont{A.~E.} \bibnamefont{Raftery}},
  \bibinfo{journal}{Journal of the American Statistical Association}
  \textbf{\bibinfo{volume}{90}}, \bibinfo{pages}{773} (\bibinfo{year}{1995}).

\bibitem[{\citenamefont{Lunn}(2013)}]{LunnDavid2013TBb:}
\bibinfo{author}{\bibfnamefont{D.}~\bibnamefont{Lunn}},
  \emph{\bibinfo{title}{The BUGS book: \\ a practical introduction to Bayesian
  analysis}}, Texts in statistical science (\bibinfo{year}{2013}), ISBN
  \bibinfo{isbn}{9781466586666}.

\bibitem[{\citenamefont{Anagnostopoulos
  et~al.}(2019)\citenamefont{Anagnostopoulos, Basilakos, and
  Saridakis}}]{Anagnostopoulos:2019miu}
\bibinfo{author}{\bibfnamefont{F.~K.} \bibnamefont{Anagnostopoulos}},
  \bibinfo{author}{\bibfnamefont{S.}~\bibnamefont{Basilakos}},
  \bibnamefont{and} \bibinfo{author}{\bibfnamefont{E.~N.}
  \bibnamefont{Saridakis}}, \bibinfo{journal}{Phys. Rev. D}
  \textbf{\bibinfo{volume}{100}}, \bibinfo{pages}{083517}
  (\bibinfo{year}{2019}), \eprint{1907.07533}.

\bibitem[{\citenamefont{Caprini and Figueroa}(2018)}]{Caprini:2018mtu}
\bibinfo{author}{\bibfnamefont{C.}~\bibnamefont{Caprini}} \bibnamefont{and}
  \bibinfo{author}{\bibfnamefont{D.~G.} \bibnamefont{Figueroa}},
  \bibinfo{journal}{Class. Quant. Grav.} \textbf{\bibinfo{volume}{35}},
  \bibinfo{pages}{163001} (\bibinfo{year}{2018}), \eprint{1801.04268}.

\bibitem[{\citenamefont{Pieroni and Barausse}(2020)}]{Pieroni:2020rob}
\bibinfo{author}{\bibfnamefont{M.}~\bibnamefont{Pieroni}} \bibnamefont{and}
  \bibinfo{author}{\bibfnamefont{E.}~\bibnamefont{Barausse}},
  \bibinfo{journal}{JCAP} \textbf{\bibinfo{volume}{07}}, \bibinfo{pages}{021}
  (\bibinfo{year}{2020}), \bibinfo{note}{[Erratum: JCAP 09, E01 (2020)]},
  \eprint{2004.01135}.

\bibitem[{\citenamefont{Babak}(2017)}]{Babak_2017}
\bibinfo{author}{\bibfnamefont{S.}~\bibnamefont{Babak}},
  \bibinfo{journal}{Journal of Physics: Conference Series}
  \textbf{\bibinfo{volume}{840}}, \bibinfo{pages}{012026}
  (\bibinfo{year}{2017}),
  \urlprefix\url{https://doi.org/10.1088/1742-6596/840/1/012026}.

\end{thebibliography}
%

\end{document}